\documentclass[preprint,preprintnumbers]{revtex4}

\usepackage{graphicx}

\usepackage{amssymb}

\usepackage{mathrsfs}

\begin{document}

\noindent

\title{Superconductivity  in Graphene  Induced by the Rotated   Layer  }

\author{D. Schmeltzer}

\affiliation{Physics Department, City College of the City University of New York,  
New York, New York 10031, USA}

 \begin{abstract}
Recent discoveries in graphene  bilayers revealed that when  one of the   layers is rotated,  superconductivity emerges. We  provide an explanation  
for this phenomenon .
We find  that due  to the layer rotations, the spinors are modified in  such way  that a repulsive interaction, becomes attractive in certain directions.
 This result is  obtained following  a sequence of steps: when  layer $2$ is        rotated  by   an  angle $\theta$ ,this rotation is equivalent to a rotation of an angle $-\theta$  of the linear  momentum .Due to the discreet lattice,   in layer $1$,  the Fourier  transform  conserves the linear  momentum $modulo$   the hexagonal reciprocal lattice vector . In layer $2$, due to the rotation,  the  linear  momentum is conserved $modulo$  the $Moire$  reciprocal lattice vector . Periodicity  is achieved  at the  $magical $ angles obtained  from the condition of commensuration of the two lattices.
We find that  the rotations  transform  the spinors  around  the nodal points, such that a repulsive interaction    becomes attractive,  giving rise to superconductivity.






\end{abstract}

\maketitle

\vspace{0.2 in}

\textbf{I. INTRODUCTION}

 A commensurate triangular $ Moire$ pattern is formed when a top-layer grapene is rotated with respect to the bottom layer  at  certain angles \cite{Crespi,Morell,Mac1}. A model  with a large amount of atoms in a commensurable unit cell was considered in order to explain the formation of the flat bands  that  might lead to superconductivity.The model consists of interlayer interaction of $\pi$ orbitals . The fit of the tight binding model was reproduced using a  Density Functional Theory  ( $DFT$) calculation of stacked bilayers.
Further progress has been achieved by \cite{Mac1,Mac2,Sharma} who computed the spectrum    considering  Coulombic interactions and phonon-mediated superconductivity.
Moire insulators have been viewed as a  surface for  a Symetric Protected Topological  phases \cite {Cenke} and by    proximity to Mott insulators  \cite{Senthil}.The proximity to Wigner crystallization  and Mott insulation  has also  been considered \cite{Philips}.The  effect  of Van Hove singularity was used   by \cite{Betouras} in analogy with high $ T_{c}$  superconductivity  calculation were  double logarithmic singularity  was the cause of Superconductivity for bar-repulsive interactions .
 The uniform rotation   generates   flat bands.  When the rotation   angle generates a periodic  $Moire$   structure commensurate with the honeycomb lattice , and the unit cell contains a large number of atoms, the Brillouine Zone
($BZ$) becomes small, the fermi 
velocity vanishes and a   flat band appears  \cite{vafek,koshino}.
\noindent 
For simplicity,  we consider the rotation  of   layer $2$ in such a way that the  site occupied by atom $B'$ is situated directly opposite from  atom $A$ in layer $1$ .
 A commensurate structure is obtained if  atom  $B'$ 
is moved by the rotation to a position formerly occupied by an atom of the same kind \cite{Lopes}.  Following  \cite{Shallcross,Lopes},we  determine  the condition for the angles $\theta(n)$ of a commensurate rotation.  
 Our goal is to investigate  the uniform rotation which can shine light on the mechanism  which is responsible for the   attraction and for  causing     \textbf{superconductivity}.

\noindent
To achive this goa,l we need to compute the effect of rotation on the spinors.
 We will   use the  tight binding model for bilayer  graphene \cite{Castro,Jackiw}  and     take into consideration  the discreetness of   the lattice \cite{Marder}.
 A two-dimensional honeycomb array of carbon atoms forming a hexagonal lattice can be viewed  as a superposition of two triangular  sublattices ,$A $ and $B$.The generators of lattice $ A$  are vectors $\vec{a}_{1}$ and  $\vec{a}_{2}$.We have three vectors $\delta_{r}$ connecting any site from lattice $A$  to nearest neighbor 
sites belonging to $B$( layer $1$).
Layer  $ i=1$  is not   rotated  and the sum over the  position of atom $ A$  and $B$  give  rise  to a summation over the reciprocal lattice vectors $\sum_{n,m}e^{i\vec{p}\cdot\vec{R}_{n,m} }=N \delta _{k_{x},G^{(1)}_{x}+G^{(2)}_{x}}\delta _{k_{y},G^{(1)}_{y}+ G^{(2)}_{y}}$ where, $\vec{R}_{n,m}= n \vec{a}_{(1)}+m \vec{a}_{(2)}$ with integers $n$ and $m$ .

 Layer $2$  is  uniformly  rotated. 
In real space the uniform rotation of the coordinates  by an angle $ \theta$
 is equivalent to a rotation $-\theta$  of  the  momentum  space $e^{i\vec {p}[-\theta]\cdot\vec{R}_{n,m} } =e^{i\vec {p}\cdot\vec{R'}_{n,m} }$  where $ \vec{R'}_{n,m}$ is the rotated vector .Performing the dicreete summation in layer $2$, we obtain 
$\sum_{n,m}e^{i\vec {p}[-\theta]\cdot\vec{R}_{n,m} }=N\delta_{k_{x},g_{x}[\theta]}\delta_{k_{y},g_{y}[\theta]}$ where $ g_{x}[\theta]=(G^{(1)}_{x}+G^{(2)}_{x})Cos[\theta]$ and $g_{y}[\theta]=(G^{(1)}_{x}+G^{(2)}_{x})Sin[\theta]$ are the reciprocal  lattice vectors  which at the magical angle  become the new reciprocal lattice vector which emerges in the following way:
layer $2$ is rotated about a site occupied by atom $B'$ directly opposite   of an atom $A$ (layer $1$). A commensurate structure is obtained if a $ B'$ atom  is moved by rotation to a  position  formerly occupied by an atom   $B '$ of the same kind.
The $Moire$ pattern  is  periodic  and a translation from the center  to the  position 
of $B'$ is a translation symmetry given by,
$cos[\theta_{i}]=\frac{3i^2+3i+0.5}{3i^2+3i+1}$,i=0,1,2..vectors. 
A superlattice  with   basis vectors $\vec{t}_{1}=i\vec{a}_{1}+(i+1)\vec{a}_{2}$;
$\vec{t}_{2}=-(i+1)\vec{a}_{1}+(2i+1)\vec{a}_{2}$ is formed  \cite{Shallcross}.
   The $Moire$  reciprocal lattice vector is  given by $g_{x}[\theta]=g_{x}[\theta_{i}]$ and $g_{y}[\theta]=g_{y}[\theta_{i}]$ ($cos[\theta_{i}]=\frac{3i^2+3i+0.5}{3i^2+3i+1}$).
 Using  the  tunneling between the layers   at  magic angles we obtain  flat bands.

 \noindent
We linearize the bilayer Hamiltonian with respect to  the nodal position and obtain  a Dirac representation .For layer $2$, the nodal position depends on the rotated angles
$\vec{K}_{2}=\frac{4\pi}{3a}\Big[ Sin[\theta],\frac{1}{\sqrt{3}}Cos[\theta]\Big]$ and $\vec{K'}_{2}=-\frac{4\pi}{3a}\Big[ Sin[\theta],\frac{1}{\sqrt{3}}Cos[\theta]\Big]$.

As a result, the  spinor will  depend  on the rotated  angles. For certain angles and for   certain  valley components, the repulsive  interaction becomes attractive:
\begin{eqnarray}
&&L_{int.}=|U[\bar{\theta}]|\int\,dy\Big[\tilde{C}^{\dagger}_{2,R,\uparrow}(p_{x}=0,y)\tilde{C}^{\dagger}_{2,R,\downarrow}(p_{x}=0,y)\tilde{C}_{2,L,\downarrow}(p_{x}=0,y)\tilde{C}_{2,L,\uparrow}(p_{x}=0,y)+\nonumber\\&&\tilde{C}^{\dagger}_{2,L,\uparrow}(p_{x}=0,y)\tilde{C}^{\dagger}_{2,L,\downarrow}(p_{x}=0,y)\tilde{C}_{2,R,\downarrow}(p_{x}=0,y)\tilde{C}_{2,R,\uparrow}(p_{x}=0,y)\Big],\nonumber\\&&
\end{eqnarray}
where $p_{x}=0$ corresponds to the nodal component $ K_{2,x}$.As a result, 
the superconductor is one dimensional with  periodicity  in the transversal direction.

The  band   satisfy $E(\vec{p})=\epsilon(\vec{p})-\mu<\Lambda$(cut-off).In our case 
the flat band, could be of the order of the chemical potential , and the condition $E(\vec{p})=\epsilon(\vec{p})-\mu<\Lambda$  is not  obeyed and  superconducting  is  not achieved  .
 
The outline of this paper is: in chapters  II and III  we consider the model
in the  real space representation.In chapter IV we linearize the model around the nodal points  obtaining a Dirac representation   for the two valleys.  We show that at  the magic angles,  the low energy bands are flat and  the spinors  transform  the repulsive    interaction to   an attractive interactions  in  certain directions. In chapter  V, we include the  spin degrees of freedom and double the number components  of  spinor.  

\vspace{0.2 in}

\textbf{III- The  real space approach}

\vspace{0.2 in}

In order to investigate the effect   of the rotation in real space,   
 we introduce the spinors $ \Psi_{i}=[a_{i},b_{i}]$,  where $a_{i}$ and  $ b_{i}$ represent the two honeycomb lattices     
   and  $i=1,2$ is the index of the two layers. 
For  layer $1$,  we have 
a  two dimensional honeycomb array of Carbon atoms forming a hexagonal lattice which can be viewed  as a superposition of two triangular  sublattices ,$A $ and $B$.The generators of lattice $ A$  are vectors $\vec{a}_{1}$ and  $\vec{a}_{2}$.We have three vectors $\delta_{r}$ connecting any site from lattice $A$  to nearest neighbor 
sites belonging to $B$.We have the representation:
\begin{eqnarray}
 &&a_{1}(\vec{R})=\frac{1}{\sqrt{L^{2}}}\sum_{\vec{k}}a_{1}(\vec{k})e^{i\vec{k}\cdot\vec{R}}\nonumber\\&&
b_{1}(\vec{R}+\vec{\delta}_{r})=\frac{1}{\sqrt{L^{2}}}\sum_{\vec{k}}b_{1}(\vec{k})e^{i\vec{k}\cdot(\vec{R}+\vec{\delta}_{r})} ;r=1,2,3
\nonumber\\&&
\end{eqnarray}
where
$a_{1}(\vec{R})=a_{1}(\vec{R}+L)$ and $b_{1}(\vec{R})=b_{1}(\vec{R}+L)$ obey the  Born-Von Karman  condition\cite{Marder} 
$\vec{k}=\frac{2\pi}{L} \vec{n}$.

\noindent
  The Hamiltonian for  layer $1$ is given by  \cite{Wallace,Jackiw}.  
\begin{eqnarray}
&&H_{1}=-\sum_{\vec{R}}\sum_{r=1,2,3}\Big[\gamma_{0}a^{\dagger}_{1}(\vec{R})b_{1}(\vec{R}+\vec{\delta_{r}})+H.C.\Big]\nonumber\\&&=-\frac{1}{N_{1}N_{2}}\sum_{\vec{R}}\sum_{r=1,2,3}\sum_{k}\sum_{p}\Big[\gamma_{0}a^{\dagger}_{1}(\vec{k})b_{1}(\vec{p})e^{-i(\vec{k}-\vec{p})\cdot \vec{R}}e^{i\vec{p}\cdot\vec{\delta}_{r}}+H.C.\Big]\nonumber\\&&
 =-\sum_{\vec{k}}\sum_{\vec{p}}\Big[b^{\dagger}_{1}(\vec{k})a_{1}(\vec{p})\phi_{1}(\vec{p})  \delta _{k_{x},p_{x}+G^{(1)}_{x}+G^{(2)}_{x}}\delta _{k_{y},p_{y}+G^{(1)}_{y}+G^{(2)}_{y}}+H.C.\Big]\nonumber\\&&
\end{eqnarray}
 Using the periodicity of the reciprocal lattice $a_{1}(p_{x},p_{y})=a_{1}(p_{x}+G^{(1)}_{x}+G^{(2)}_{x},p_{y}+G^{(1)}_{y}+G^{(2)}_{y})$,and $b_{1}(p_{x},p_{y})=b_{1}(p_{x}+G^{(1)}_{x}+G^{(2)}_{x},p_{y}++G^{(1)}_{y}+G^{(2)}_{y})$ we obtain:
\begin{equation}
H_{1}=\sum_{\vec{p}\in  B.Z.}\Big[a^{\dagger}_{1}(\vec{p})b_{1}(\vec{p})\phi_{1}(\vec{p})  +H.C.\Big]
\label{H1}
\end{equation}
where the  hexagonal  reciprocal  lattice vectors are $\vec{G}^{(1)}=\frac{2\pi}{3a}\Big[1,\sqrt{3}\Big]$ and $\vec{G}^{(2)}=\frac{2\pi}{3a}\Big[1,-\sqrt{3}\Big]$,   with the two   Bravais unit cell vectors $ \vec{a}_{(1)}=\frac{a}{2}\Big[3,
\sqrt{3}\Big]$, $ \vec{a}_{(2)}=\frac{a}{2}\Big[3,
-\sqrt{3}\Big]$. 
 
\noindent
The discrete sum  over the integers $n$ and $m$ ,$\vec{R}_{n,m}= n \vec{a}_{(1)}+m \vec{a}_{(2)}$ determines the position of the lattice atom    $A$  \cite{Castro}. 
 Atom $B$ is  given by the relative vectors  $\vec{\delta_{r}}$ ,r=1,2,3. with respect to  atom $A$  at position   $\vec{R}_{n,m}$.The sum over the vectors $\vec{\delta}_{r}$ determines the function  $\phi_{1}(\vec{p})$.
\begin{equation}
	 \phi_{1}(\vec{p})=-\gamma_{0}\sum_{r=1,2,3}e^{-i \vec{p}\cdot\vec{\delta_{r}}}=e^{-ip_{x}}\Big[1+2e^{i\frac{3}{2}p_{x}}Cos[\frac{\sqrt{3}}{2}p_{y}\Big]
\label{phi1}
\end{equation}
The location of the two   nodes in layer $i=1$ is given by $\vec{K}_{1}=[0,\frac{4\pi}{3\sqrt{3}}]
$ and $\vec{K'}_{1}=[0,\frac{-4\pi}{3\sqrt{3}}]$ which obey 
$\phi_{1}(\vec{p}=\vec{K}_{1})=0$ and $\phi_{1}(p=\vec{K'}_{1})=0$.
\noindent
 Layer $2$  is rotated with respect to  layer  $1$ . In layer $1$ the atoms are    $A_{1}$ and $B_{1}$ while   in layer $2$  the atoms  are  $A'=A_{2} $ and $B'=B_{2}$  . In a stacked $A,B$ bilaye,r  $A'$ and $B'$   have  the same horizontal position as atoms $A_{1}$ and $B_{1}$.

The rotation of layer $2$ occurs in such   a way that the  site occupied by atom $B'$ is located directly opposite  an atom $A$ (layer 1).A commensurate structure is obtained if a $ B'$ atom  is moved by rotation to a position  formerly occupied by an atom $B'$ of the same kind.
The $Moire$ pattern  is  periodic  and a translation from the center  to the  position 
of $B'$ is a translation symmetry.

$cos[\theta_{i}]=\frac{3i^2+3i+0.5}{3i^2+3i+1}$,where the vector  $i$  is i=0,1,2... 

The superlattice bases are  $\vec{t}_{1}=i\vec{a}_{1}+(i+1)\vec{a}_{2}$;
$\vec{t}_{2}=-(i+1)\vec{a}_{1}+(2i+1)\vec{a}_{2}$;

 In addition, in layer $2$,  the vector $\vec{R}_{n,m}$  is replaced  by the rotated vector $\vec{R'}_{n,m}$:
\noindent 
$ \vec{R'}_{n,m}=\vec{R'}(\theta)_{n,m}=\Big[R'_{x},R'_{y}\Big]_{n,m}=\Big[R_{x}Cos[\theta]- R_{y}Sin[\theta],R_{x}Sin[\theta]+R_{y}Cos[\theta]\Big]_{n,m} $ 
\noindent
and  
$\vec{\delta}_{r}$ is replaced by, 
\noindent
$\vec{\delta'}_{r}(\theta)=\Big[\delta'_{r,x},\delta'_{r,y}\Big]=\Big[\delta_{r,x}Cos[\theta]- \delta_{r,y}Sin[\theta],\delta_{r,x}Sin[\theta]+\delta_{r,y}Cos[\theta]\Big]$
\noindent
  The Hamiltonian for  the rotated  layer $2$ is given by:
\begin{eqnarray}
&&H_{2}=-\sum_{\vec{R}}\sum_{r=1,2,3}\gamma_{0}\Big[ b{\dagger}_{2}(\vec{R'})a_{2}(\vec{R'}-\vec{\delta}_{r'})+H.C.\Big]\nonumber\\&&
 =  -\frac{1}{N_{1}N_{2}}\sum_{\vec{R}}\sum_{r=1,2,3}\sum_{\vec{k}}\sum_{\vec{p}}\Big[\gamma_{0}b^{\dagger}_{2}(\vec{k})a_{2}(\vec{p})e^{-i(\vec{k}-\vec{p})\cdot \vec{R}'}e^{-i\vec{p}\cdot\vec{\delta}_{r'}}+H.C.\Big]\nonumber\\&&
= -\frac{1}{N_{1}N_{2}}\sum_{\vec{R}}\sum_{r=1,2,3}\sum_{\vec{k}}\sum_{\vec{p}}\Big[\gamma_{0}b^{\dagger}_{2}(\vec{k})a_{2}(\vec{p})e^{-i(\vec{k}[-\theta]-\vec{p}[-\theta])\cdot \vec{R}}e^{-i\vec{p}[-\theta]\cdot\vec{\delta}_{r}}+H.C.\Big]\nonumber\\&&
\end{eqnarray}
When we rotate the coordinates of layer $2$ by an angle $\theta$, the momentum is rotated by angle $-\theta$,
 $\vec{k}[-\theta] =\Big[k_{x}Cos[\theta]+k_{y}Sin[\theta],k_{y}Cos[\theta]-k_{x}Sin[\theta]\Big]$
\begin{eqnarray}
&&H_{2}=-\sum_{\vec{k}}\sum_{\vec{p}}\Big[a_{2}(\vec{k})b_{2}(\vec{p})\varphi_{2}(\vec{p}) \delta _{k_{x},p_{x}+g_{x}[\theta] }\delta _{k_{y},p_{y}+g_{y}[\theta]}+H.C.\Big]\nonumber\\&&
g_{x}[\theta]=(G^{(1)}_{x}+G^{(2)}_{x})Cos[\theta]; \hspace{0.2 in}
g_{y}[\theta]=(G^{(1)}_{x}+G^{(2)}_{x})Sin[\theta]
\end{eqnarray}
$g_{x}[\theta]$ and $g_{y}[\theta]$ are the new BZ with two lattice vectors $L_{1}$,$L_{2}$.

\noindent
At special angles $ g_{x}[\theta=\theta_ {magic}]=g^{Moire}_{x}$, $ g_{y}[\theta=\theta _{magic}]=g^{Moire}_{y}$ given by $ Cos[\theta(n)]=\frac{3n^2+3n +\frac{1}{2}}{3n^2+3n+1}$ n=1,2...\cite{Shallcross,Lopes}, we obtain the Moire comensurate rotations such that  $g_{x}[\theta_ {magic}=\theta(n)]L_{1}=2\pi$ ,$g_{y}[\theta_ {magic}=\theta(n)]L_{2}=2\pi$.

\noindent
 Using the  periodicity in the BZ with respect  to the  $Moire$ reciprocal lattice $a_{2}(p_{x},p_{y})=a_{2}(p_{x}+g_{x}[\theta],p_{y}+g_{y}[\theta])$ , $b_{2}(p_{x},p_{y})=b_{2}(p_{x}+g_{x}[\theta],p_{y}+g_{y}[\theta])$, we find that  at the magic angle $ [\theta _{magic}] $ it is commensurate with  the hexagonal  lattice $a_{2}(p_{x},p_{y})=a_{2}(p_{x}+g_{x}[\theta=0],p_{y}+g_{y}[\theta]=0)$ , $b_{2}(p_{x},p_{y})=b_{2}(p_{x}+g_{x}[\theta=0],p_{y}+g_{y}[\theta=0])$ .As a result of the periodicity and commensuration we obtain:
\begin{equation}
H_{2}=\sum_{\vec{p}\in   Moire B.Z}\Big[b_{2}(\vec{p})a_{2}(\vec{p})\varphi_{2}(\vec{p})+H.C.\Big]
\label {2layer}
\end{equation}
We replace $\vec{p}\cdot\vec{\delta'}_{r}(\theta)$ with  $\vec{p}[-
\theta]\cdot\vec{\delta}_{r}$ and compute  $\varphi_{2}(p)$:
\begin{eqnarray}
  &&\varphi_{2}(p)=\phi_{1}(\vec{p}[-\theta]) 
=\gamma_{0}\sum_{r=1,2,3}e^{-i\vec{p}\cdot\vec{\delta'}_{r}} =\gamma_{0}\sum_{r=1,2,3}e^{-i\vec{p[-\theta])}\cdot\vec{\delta}_{r}}
\nonumber\\&&
 \varphi_{2}(\vec{p})=\gamma_{0}e^{-ip_{x}Cos[\theta]} e^{-ip_{y}Sin[\theta]}\Big[1+ 2e^{i\frac{3}{2}p_{x}Cos[\theta]}e^{i\frac{3}{2}p_{y}Sin[\theta]}Cos[\frac{\sqrt{3}}{2}(p_{x}Cos[\theta]-p_{y}Sin[\theta])]\Big]\nonumber\\&&
\end{eqnarray}
The Dirac points in   layer $2$  are at  momentum $\vec{K}_{2}=\frac{4\pi}{3a}\Big[Sin[\theta], \frac{1}{\sqrt{3}}Cos[\theta]\Big] $   and $\vec{K'}_{2}=-\frac{4\pi}{3a}\Big[Sin[\theta], \frac{1}{\sqrt{3}}Cos[\theta]\Big] $ .

The tunneling Hamiltonian between layer $1$  and $2$ is given by \cite{Lopes}:
\begin{eqnarray}
&&H_{\perp,3}=-\frac{\gamma_{3}}{\gamma_{0}}\sum_{r=1,2,3}\sum_{\vec{R}}\Big[b^{\dagger}_{2}(\vec{R'})a_{1}(\vec{R})+a^{\dagger}_{2}(\vec{R'})b_{1}(\vec{R}+\vec{\delta}_{r})\Big]=\nonumber\\&&
-\frac{1}{N_{1}N_{2}}\frac{\gamma_{3}}{\gamma_{0}}\sum_{r=1,2,3}\sum_{\vec{R}}\sum_{\vec{k}}\sum_{\vec{p}}\Big[b^{\dagger}_{2}(\vec{k})a_{1}(\vec{p})e^{-i(\vec{k}[-\theta]-\vec{p})\cdot\vec{R}}e^{-ik[-\theta]\cdot\vec{\delta}_{r}}+a_{2}(\vec{k})b_{1}(\vec{p})e^{-i(\vec{k}[-\theta]-\vec{p})\cdot\vec{R}}e^{ip\cdot\vec{\delta}_{r}}\Big]\nonumber\\&&
-\frac{1}{N_{1}N_{2}}\frac{\gamma_{3}}{\gamma_{0}}\sum_{r=1,2,3}\sum_{\vec{R}}\sum_{\vec{k}}\sum_{\vec{p}}\Big[b^{\dagger}_{2}(\vec{k})a_{1}(\vec{p})e^{-i(\vec{k}[-\theta]-\vec{p})\cdot\vec{R}}\varphi^{*}_{2}(\vec{k})+a_{2}(\vec{k})b_{1}(\vec{p})e^{-i(\vec{k}[-\theta]-\vec{p})\cdot\vec{R}} \phi_{1}(\vec{p})\Big]=\nonumber\\&&
-\frac{\gamma_{3}}{\gamma_{0}}\sum_{\vec{k}}\sum_{\vec{p}}\Big[b^{\dagger}_{2}(\vec{k})a_{1}(\vec{p})\delta_{k_{x}, p_{x}Cos[\theta]-p_{y} Sin[\theta]+g_{x}[\theta]} \delta_{k_{y}, p_{y}Cos[\theta]+p_{x} Sin[\theta]+g_{y}[\theta]}\varphi^{*}_{2}(\vec{k})\nonumber\\&&+a_{2}(\vec{k})b_{1}(\vec{p}) \delta_{k_{x}, p_{x}Cos[\theta]-p_{y} Sin[\theta]+g_{x}[\theta]} \delta_{k_{y}, p_{y}Cos[\theta]+p_{x} Sin[\theta]+g_{y}[\theta]}\phi_{1}(\vec{p})\Big],\nonumber\\&&
\end{eqnarray}
  where  $\vec{p}[-\theta]=\Big[p_{x}Cos[\theta]+p_{y}Sin[\theta],p_{y}Cos[\theta]-p_{x}Sin[\theta]\Big]$

\noindent
For small angle rotations,  we replace $H_{\perp,3}$ with the $Moire$ BZ: 
 \begin{equation}
H_{\perp,3}=-\frac{\gamma_{3}}{\gamma_{0}}\sum_{\vec{p}\in   Moire B.Z}\Big[b^{\dagger}_{2}(\vec{p})a_{1}(\vec{p})\phi_{1}(\vec{p})+a^{\dagger}_{2}(\vec{p})b_{1}(\vec{p})\phi^{*}_{1}(\vec{p})\Big]
\label{h3}
\end{equation}
Here,  the  tunneling coupling constant  is given by  $\gamma_{0}=2.8 ev$,$\gamma_{3}=0.3 ev$  \cite{Castro}.

\vspace{0.2 in} 

\textbf{IV-Computation of the eigenvalues}

\vspace{0.3 in}
We will compute the eigenvalues for small angles which correspond to a $Moire $ commensurate lattice.
The Hamiltonian $ H_{1}$ and $H_{2}$ are diagonalyzed using  the  following representation :
 for layer $ i=1$ the eigenvalues are $ \pm| \varphi_{1} (\vec{p})|$, with the   two eigenvectors  $ u_{1}(\vec{p})=\frac{1}{\sqrt{2}}\Big[1,-e^{-i\alpha_{1}(\vec{p})}\Big]^{
T} =\frac{1}{\sqrt{2}}\Big[1,-\frac{|\phi_{1}(\vec{p}|}{\phi_{1}(\vec{p}}\Big]^{T}$ and for negative eigenvalues  the eigenvector  is $v_{1}(\vec{p})=\frac{1}{\sqrt{2}}\Big[1,e^{-i\alpha_{1}(\vec{p})}\Big]^{T}$    

\noindent
 For layer $i=2$ we have the eigenvalue $\pm|\varphi_{2}(p)|$
The eigenvector  for layer $ i=2 $  is   $ u_{2}(\vec{p})=\frac{1}{\sqrt{2}}\Big[1,-e^{-i\alpha_{2}(\vec{p})}\Big]^{T}=\frac{1}{\sqrt{2}}\Big[1,-\frac{\varphi^{*}_{2}(p)}{|\varphi_{2}(p)|}\Big]^{T}$ 
for positive eigenvalues, while  for the negative eigenvalues  $v_{2}(\vec{p})=\frac{1}{\sqrt{2}}\Big[1,e^{-i\alpha_{2}(\vec{p})}\Big]^{T}$
\begin{eqnarray}
&&\Psi_{2}(\vec{p})=C_{2}(\vec{p})u_{2}(\vec{p})+D^{\dagger}_{2}(\vec{p}) v_{2}(\vec{p}) \nonumber\\&& 
H_{2}=\sum_{\vec{p}}C^{\dagger}_{2}(\vec{p})C_{2}(\vec{p})|\varphi_{2}(\vec{p}))| +D^{\dagger}_{2}(\vec{p})D_{2}(\vec{p})
|\varphi_{2}(\vec{p})|\nonumber\\&&
\Psi_{1}(\vec{p})=C_{1}(\vec{p})u_{1}(\vec{p})+D^{\dagger}_{1}(\vec{p}) v_{1}(\vec{p}) \nonumber\\&& 
H_{1}=\sum_{\vec{p}}C^{\dagger}_{1}(\vec{p})C_{1}(\vec{p})\phi_{1}(\vec{p}))| +D^{\dagger}_{1}(\vec{p})D_{1}(\vec{p})
\phi_{1}(\vec{p})|\nonumber\\&&
\end{eqnarray}
$C^{\dagger}_{2}(\vec{p})$,$C_{2}(\vec{p})$,$C^{\dagger}_{1}(\vec{p})$,$C_{1}(\vec{p})$  are the particle operators and $ D^{\dagger}_{2}(\vec{p})$,$D_{2}(\vec{p})$,$D^{\dagger}_{1}(\vec{p})$,$D_{1}(\vec{p})$ are the anti-particles operators.

\noindent
\textbf{We neglect the anti -particles and rewrite the Hamiltonian  in terms of the particle operators only.}
\noindent
We consider the small angles such that the   $Moire$ lattice is commensurate with the hexagonal  lattice.

\begin{eqnarray}
&&H_{1}=\sum_{\vec{p}}C^{\dagger}_{1}(\vec{p})C_{1}(\vec{p})|\phi_{1}(\vec{p})|\nonumber\\&&
H_{2}=\sum_{\vec{p}}C^{\dagger}_{2}(\vec{p})C_{2}(\vec{p})|\varphi_{2}(\vec{p})|\\&&\nonumber\\&&
H_{\perp,3}=-\frac{\gamma_{3}}{\gamma_{0}}\sum_{\vec{p}\in   Moire B.Z}C^{\dagger}_{2}(\vec{p})C_{1}(\vec{p})|\phi_{1}(\vec{p})|  Cos[\alpha_{2}(\vec{p})+\alpha_{1}(\vec{p})]+H.C.\nonumber\\&&
\end{eqnarray}
The lowest eigenvalue is given by:
\begin{equation}
E^{-}(\vec{p})=-\mu+\frac{1}{2}\Big( |\phi_{1}(\vec{p})|+|\varphi_{2}(\vec{p})|-\sqrt{(|\phi_{1}(\vec{p})|-|\varphi_{2}(\vec{p})|)^{2}+4((\frac{\gamma_{3}}{ \gamma_{0}}) |\phi_{1}(\vec{p})|Cos[\alpha_{2}(\vec{p})+\alpha_{1}(\vec{p})])^{2}}\Big)
\label{equation}
\end{equation}
We include   the chemical potential $\mu$ and observe that the band is quasi- flat .

In order to obtain a  better description of the bands,we will   expand  the model around the Dirac Cones and we will observe  the Dirac dispersion.

\vspace{0.2 in}

\textbf{IV-The continuum model}

\vspace{0.2 in}

In order to see how    the interactions are affected by  the rotations  we will use a  continuum model. The continuum model will show the Dirac dispersion around the Dirac cones.
 For each layer we introduce a linear model around  the position  of the two Dirac nodes.
For layer $1$, we replace $\vec{p}=\vec{K'}_{1}+\vec{q}$  for the left valley  and $\vec{p}=\vec{K}_{1}+\vec{q}$ for the right valley.  The two valley are represented by   the Pauli matrix $\vec{\tau}$.  
\begin{eqnarray}
&&\sum_{\vec{p}\in  B.Z}.\Big[a^{\dagger}_{1}(\vec{p})b_{1}(\vec{p})\phi_{1}(\vec{p})  +H.C.\Big]\nonumber\\&&
=\sum_{\vec{q}}\Big[a^{\dagger}_{1}(\vec{p}=\vec{K'}_{1}+\vec{q})b_{1}(\vec{p}=\vec{K'}_{1}+\vec{q}_{1})\phi_{1}(\vec{p}=\vec{K'}_{1}+\vec{q}_{1})  +\nonumber\\&&+a^{\dagger}_{1}(\vec{p}=\vec{K}_{1}+\vec{q})b_{1}(\vec{p}=\vec{K}_{1}+\vec{q})\phi_{1}(\vec{p}=\vec{K}_{1}+\vec{q})  +H.C.\Big]\nonumber\\&&
=-\sum_{\vec{q}}\Big[a_{1,L}^{\dagger}(\vec{q})b_{1,L}(\vec{q})\phi_{1,L}(\vec{q})  +a^{\dagger}_{1,R}(\vec{q})b_{1,R}(\vec{q})\phi_{1,R}(\vec{q})  +H.C.\Big]\nonumber\\&&
=\sum_{\vec{q}}\Big[v_{F}\Phi_{1,L}^{\dagger}(\vec{q})\Big(\tau_{1}q_{y}-\tau_{2}q_{x}\Big)\Phi_{1,L}(\vec{q})+v_{F}\Phi_{1,R}^{\dagger}(\vec{q})\Big(-\tau_{1}q_{y}-\tau_{2}q_{x}\Big)\Phi_{1,R}(\vec{q})\Big]\nonumber\\&&
\end{eqnarray}
where
\begin{eqnarray}
 &&\Phi_{1,L}(\vec{q})=\Big[a_{1}(\vec{p}=\vec{K'}_{1}+\vec{q}),b_{1}(\vec{p}=\vec{K'}_{1}+\vec{q})\Big]^{T}=\Big[a_{1,L}(\vec{q}),b_{1,L}(\vec{q})\Big]^{T}\nonumber\\&& \Phi_{1,R}(\vec{q})=\Big[a_{1}(\vec{p}=\vec{K}_{1}+\vec{q}),b_{1}(\vec{p}=\vec{K}_{1}+\vec{q})\Big]^{T}=\Big[a_{1,R}(\vec{q}),b_{1,R}(\vec{q})\Big]^{T} \nonumber\\&&
\Psi_{1}(\vec{r})=\Phi_{1,L}(\vec{r})e^{i\vec{K'}_{1}\cdot \vec{r}}+\Phi_{1,R}(\vec{r})e^{i\vec{K}_{1}\cdot \vec{r}}\nonumber\\&&
\Psi_{1}(\vec{p})=\Psi_{1}(\vec{p})\mu[-p_{x}]+\Psi_{1}(\vec{p})\mu[p_{x}]=\Phi_{1,L}(\vec{q}+\vec{K'}_{1})\mu[-q_{x}-K'_{x,1}]+\Phi_{1,R}(\vec{q}+\vec{K}_{1})\mu[q_{x}+K_{x,1}],\nonumber\\&&
\end{eqnarray}
where $\mu[q_{x}+K_{x,1}]$ is the step function which obeys $\mu[-q_{x}-K'_{x,1}]+\mu[q_{x}+K_{x,1}]=1$.
 From equation $(15)$ we obtain the eigen spinors and represent $\Phi_{1,L}(\vec{q})$, $\Phi_{1R}(\vec{q})$ in terms of the valley operators $C_{1,L}(\vec{q})$ , $C_{1,R}(\vec{q})$ 
\begin{eqnarray}
&&\Phi_{1,L}(\vec{q})=C_{1,L}(\vec{q})U_{1,L}(\vec{q})\nonumber\\&&\Phi_{1,R}(\vec{q})=C_{1,R}(\vec{q})U_{1,R}(\vec{q})\nonumber\\&&
U_{1,L}(\vec{q})=\frac{1}{\sqrt{2}}\Big[1,-ie^{\alpha(\vec{q})}\Big]^{T},\hspace{0.1in}U_{1,R}(\vec{q})=\frac{1}{\sqrt{2}}\Big[1,ie^{-i\alpha(\vec{q})}\Big]^{T},\hspace{0.05in}\alpha(\vec{q)}=ArcTan\Big(\frac{q_{y}}{q_{x}}\Big)\nonumber\\&&
\Psi_{1}(\vec{r})=\sum_{\vec{q}}\Big[C_{1,L}(\vec{q})U_{1,L}(\vec{q})e^{i(\vec{q}+\vec{K'}_{1})\cdot\vec{r}}+C_{1,R}(\vec{q})U_{1,R}(\vec{q})e^{i(\vec{q}+\vec{K}_{1})\cdot\vec{r}}\Big]\nonumber\\&&=\sum_{\vec{p}}\Big[C_{1,L}(\vec{p}-\vec{K'}_{1})U_{1,L}(\vec{p}-\vec{K'}_{1})e^{i\vec{p}\cdot\vec{r}}\mu[-p_{x}]+C_{1,R}(\vec{p})U_{1,R}(\vec{p}-\vec{K}_{1})e^{i\vec{p}\cdot\vec{r}}\mu[p_{x}],\nonumber\\&&
\tilde{C}_{1,L}(\vec{p})=C_{1,L}(\vec{p}-\vec{K'}_{1}),\hspace{0.1in}\tilde{C}_{1,R}(\vec{p})=C_{1,R}(\vec{p}-\vec{K}_{1})\nonumber\\&&
\end{eqnarray}
For  layer $i=2$  with the  rotated  nodes     at  momentum $\vec{K}_{2}=\frac{4\pi}{3a}\Big[Sin[\theta], \frac{1}{\sqrt{3}}Cos[\theta]\Big] $   and $\vec{K'}_{2}=-\frac{4\pi}{3a}\Big[Sin[\theta], \frac{1}{\sqrt{3}}Cos[\theta]\Big] $ , eq.$(8)$ gives the linearized form:
\begin{eqnarray}
&&H_{2}=\sum_{\vec{p}\in   Moire B.Z}\Big[b_{2}(\vec{p})a_{2}(\vec{p})\varphi_{2}(\vec{p})+H.C.\Big]\nonumber\\&& =\sum_{\vec{q}}\Big[b_{2}(\vec{p}=\vec{q}+ \vec{K}_{2})a_{2}(\vec{p}=\vec{q}+ \vec{K}_{2})\varphi_{2}(\vec{p}=\vec{q}+ \vec{K}_{2})\nonumber\\&&+b_{2}(\vec{p}=\vec{q}+ \vec{K'}_{2})a_{2}(\vec{p}=\vec{q}+ \vec{K'}_{2})\varphi_{2}(\vec{p}=\vec{q}+ \vec{K'}_{2})+H.C.\Big]=\nonumber\\&&-\sum_{\vec{q}}\Big[a_{2,L}^{\dagger}(\vec{q})b_{2,L}(\vec{q})\phi_{2,L}(\vec{q})  +a^{\dagger}_{2,R}(\vec{q})b_{2,R}(\vec{q})\varphi_{2,R}(\vec{q})  +H.C.\Big]\nonumber\\&&
=\sum_{\vec{q}}\Big[v_{F}\Phi_{2,L}^{\dagger}(\vec{q})\Big(\tau_{1}(q_{y}Cos[\theta]-q_{x}Sin[\theta])+\tau_{2}(q_{y}Sin[\theta]+q_{x}Cos[\theta])\Big)\Phi_{2,L}(\vec{q})\nonumber\\&&+v_{F}\Phi_{2,R}^{\dagger}(\vec{q})\Big(\tau_{1}(q_{y}Cos[\theta]-q_{x}Sin[\theta])-\tau_{2}(q_{x}Cos[\theta]+q_{y}Sin[\theta])\Big)\Phi_{2,R}(\vec{q})\Big]\nonumber\\&&
\end{eqnarray}
The field in layer $i=2$  has the representation:
\begin{eqnarray}
&&\Psi_{2}(\vec{r'})=\Phi_{2,L}(\vec{r'})e^{i\vec{K'}_{2}\cdot \vec{r}}+\Phi_{2,R}(\vec{r'})e^{i\vec{K}_{2}\cdot \vec{r}}\nonumber\\&&
\end{eqnarray}
with the representation : 
  $\Phi_{2,L}(\vec{q})$, $\Phi_{2,R}(\vec{q})$ in terms of the operators $C_{2,L}(\vec{q}$ , $C_{2,R}(\vec{q})$ 
\begin{eqnarray}
&&\Phi_{2,L}(\vec{q})=C_{2,L}(\vec{q})U_{2,L}(\vec{q}),\hspace{0.1in}\Phi_{2,R}(\vec{q})=C_{2,R}(\vec{q})U_{2,R}(\vec{q}\nonumber\\&&
U_{2,L}=\frac{1}{\sqrt{2}}\Big[1,i e^{i[\theta]}e^{-i\alpha(\vec{q})}\Big]^{T},\hspace{0.1in}U_{2,R}=\frac{1}{\sqrt{2}}\Big[1,-i  e^{-i[\theta]}e^{i\alpha(\vec{q})}\Big]^{T},\hspace{0.05in}\alpha(\vec{q)}=ArcTan\Big(\frac{q_{y}}{q_{x}}\Big)\nonumber\\&&
\Psi_{2}(\vec{r'})=\sum_{\vec{q}}\Big[C_{2,L}(\vec{q})U_{2,L}(\vec{q})e^{i(\vec{q}[-\theta]-\vec{K'}_{2})\cdot\vec{r}}+C_{2,R}(\vec{q})U_{2,R}(\vec{q})e^{i(\vec{q}[-\theta]-\vec{K}_{2})\cdot\vec{r}}\Big]\nonumber\\&&=\sum_{\vec{p}}\Big[\tilde{C}_{2,L}(\vec{p})U_{2,L}(\vec{p}-\vec{K'}_{2})e^{i\vec{p}[-\theta]\cdot\vec{r}}+\tilde{C}_{2,R}(\vec{p})U_{2,R}(\vec{p}-\vec{K}_{2})e^{i\vec{p}[-\theta]\cdot\vec{r}}\nonumber\\&&
\end{eqnarray} 
The rotated  layer $i=2$ ,$\Psi_{2}(\vec{r'})$ is represented in terms of the rotated 
  Dirac
nodes ,  $\vec{K'}_{2}=-\frac{2\pi}{3a}\Big[Sin[\theta],\frac{1}{\sqrt{3}}Cos[\theta]\Big]$, $\vec{K}_{2}=\frac{2\pi}{3a}\Big[Sin[\theta],\frac{1}{\sqrt{3}}Cos[\theta]\Big]$ .

\noindent
The eigenvalues for the two  layers  around the two Dirac points for particles  with respect to  the momentum  $\vec{q}$ are :
 $\epsilon_{1,L}(\vec{q})=\epsilon_{1,R}(\vec{q})=v_{F}\sqrt{(q_{x})^2+(q_{y})^2}$
 and 
$\epsilon_{2,L}(\vec{q})=\epsilon_{2,R}(\vec{q})=v_{F}\sqrt{(q_{x})^2+(q_{y})^2}$.

\noindent
 For  the  tunneling Hamiltonian, we obtain:  
\begin{eqnarray}
&&H_{\perp,3}=-\gamma_{3}\sum_{\vec{R}}\sum_{r=1,2,3}\Big[ b^{\dagger}_{2}(\vec{R'}-\vec{\delta}_{r'})a_{1}(\vec{R})+a^{\dagger}_{2}(\vec{R'})b_{1}(\vec{R}+\vec{\delta}_{r})+H.C.\Big]\nonumber\\&&=-\gamma_{3}\sum_{\vec{R}}\sum_{\vec{\delta}_{r}}
\Big[(\Phi^{\dagger}_{2}(\vec{R'}-\vec{\delta}_{r'})_{b}(\Phi_{1}(\vec{R}))_{a}+(\Phi^{\dagger}_{2}(\vec{R'}))_{a}(\Phi_{1}(\vec{R}+\vec{\delta}_{r}))_{b}+H.C.\Big]\nonumber\\&&
\end{eqnarray}
where 
\begin{eqnarray}
&&\Phi_{2}(\vec{R'}-\vec{\delta}_{r'})_{b}=\sum_{\vec{p}}\Big[\tilde{C}_{2,L}(\vec{p})U^{(2)}_{2,L}(\vec{p}-\vec{K'}_{2})e^{i\vec{p}[-\theta]\cdot(\vec{R}-\vec{\delta}_{r})}+\tilde{C}_{2,R}(\vec{p})U^{(2)}_{2,R}(\vec{p}-\vec{K}_{2})e^{i\vec{p}[-\theta]\cdot(\vec{R}-\vec{\delta}_{r})}\Big]\nonumber\\&&
\Phi_{1}(\vec{R})_{a}=\sum_{\vec{p}}\Big(\tilde{C}_{1,L}(\vec{p})U^{(1)}_{1,L}(\vec{p}-\vec{K'}_{1})e^{i\vec{p}\cdot\vec{R}}+\tilde{C}_{1,R}(\vec{p})U^{(1)}_{1,R}(\vec{p}-\vec{K}_{1})e^{i\vec{p}\cdot\vec{R}}\Big)\nonumber\\&&
\end{eqnarray}
\begin{eqnarray}
	&&H_{\perp,3}=-\gamma_{3}\sum_{\vec{p'}}\sum_{\vec{p}}\sum_{\vec{R}}\tilde{C}_{2,L}(\vec{p})U^{(2)}_{2,L}(\vec{p}-\vec{K'}_{2})e^{i\vec{p}[-\theta]\cdot\vec{R}}\sum_{1,2,3}e^{-i\vec{p}[-\theta]\cdot\vec{\delta}_{r}}+\nonumber\\&&\tilde{C}_{2,R}(\vec{p})U^{(2)}_{2,R}(\vec{p}-\vec{K}_{2})e^{i\vec{p}[-\theta]\cdot\vec{R}}\sum_{1,2,3}e^{-i\vec{p}[-\theta]\cdot\vec{\delta}_{r}}\Big)^{\dagger}\cdot\Big(\tilde{C}_{1,L}(\vec{p'})U^{(1)}_{1,L}(\vec{p}-\vec{K'}_{1})e^{i\vec{p}\cdot\vec{R}}+\tilde{C}_{2,R}(\vec{p'})U^{(1)}_{2,R}(\vec{p}-\vec{K}_{2})e^{i\vec{p}\cdot\vec{R}}\Big)\nonumber\\&&+\Big(\tilde{C}_{2,L}(\vec{p})U^{(1)}_{2,L}(\vec{p}-\vec{K'}_{2})e^{i\vec{p}[-\theta]\cdot\vec{R}}+\tilde{C}_{2,R}(\vec{p})U^{(1)}_{2,R}(\vec{p}+\vec{K}_{2})e^{i\vec{p}[-\theta]\cdot\vec{R}}\Big)^{\dagger}\cdot\nonumber\\&&\Big(\tilde{C}_{1,L}(\vec{p'})U^{(2)}_{1,L}(\vec{p}-\vec{K'}_{1})e^{i\vec{p}\cdot\vec{R}}\sum_{1,2,3}e^{-i\vec{p}\cdot\vec{\delta}_{r}}-\tilde{C}_{2,R}(\vec{p'})U^{(2)}_{2,R}(\vec{p}-\vec{K}_{2})e^{i\vec{p}\cdot\vec{R}}\sum_{1,2,3}e^{-i\vec{p}\cdot\vec{\delta}_{r}}\Big)+H.C.\Big]\nonumber\\&&
\end{eqnarray}
Using the periodicity with respect to the $Moire$ reciprocal lattice which at magical  angles is commensurate with the  honeycomb lattice, at  small angles we obtain in the  $Moire$ BZ  the  representation:
\begin{eqnarray}
&&H_{\perp,3}\approx-\frac{\gamma_{3}}{\gamma_{0}}\sum_{\vec{p}\in   Moire B.Z}\Big[\Big(\tilde{C}_{2,L}(\vec{p})U^{(2)}_{2,L}(\vec{p}-\vec{K'}_{2})\phi_{1}(\vec{p})+\tilde{C}_{2,R}(\vec{p})U^{(2)}_{2,R}(\vec{p}-\vec{K}_{2})\phi_{1}(\vec{p})\Big)^{\dagger}\cdot\nonumber\\&&\Big(\tilde{C}_{1,L}(\vec{p})U^{(1)}_{1,L}(\vec{p}-\vec{K'}_{1})+\tilde{C}_{2,R}(\vec{p})U^{(1)}_{2,R}(\vec{p}-\vec{K}_{2})\Big)+\nonumber\\&&\Big(\tilde{C}_{2,L}(\vec{p})U^{(1)}_{2,L}(\vec{p}-\vec{K'}_{2})+\tilde{C}_{2,R}(\vec{p})U^{(1)}_{2,R}(\vec{p}-\vec{K}_{2})\Big)^{\dagger}\Big(\tilde{C}_{1,L}(\vec{p'})U^{(2)}_{1,L}(\vec{p}-\vec{K'}_{1})\phi_{1}(\vec{p})+\tilde{C}_{2,R}(\vec{p})U^{(2)}_{2,R}(\vec{p}-\vec{K}_{2})\nonumber\\&&\phi_{1}(\vec{p})\Big)+H.C.\Big]\nonumber\\&&
\end{eqnarray}

\begin{eqnarray}
&&H=H_{L}+H_{R}+H_{L,R}\nonumber\\&&
H_{L}=\sum_{\vec{p}}\Big[\Big(\tilde{C}^{\dagger}_{1,L}(\vec{p})\tilde{C}(\vec{p})_{1,L}\epsilon_{2,L}(\vec{p}-\vec{K'}_{1})+\tilde{C}^{\dagger}_{2,L}(\vec{p})\tilde{C}(\vec{p})_{2,L}\epsilon_{2,L}(\vec{p}-\vec{K'}_{2})\Big)+\nonumber\\&&g_{1L,2L}\tilde{C}^{\dagger}_{1,L}(\vec{p}) \tilde{C}_{2,L}(\vec{p})+H.C.\Big]
\nonumber\\&&H_{R}=\Big[\tilde{C}^{\dagger}_{1,R}(\vec{p})\tilde{C}_{1,R}(\vec{p})\epsilon_{1,R}(\vec{p}-\vec{K}_{1})+\tilde{C}^{\dagger}_{2,R}(\vec{p})\tilde{C}_{2,R}(\vec{p})\epsilon_{2,R}(\vec{p}-\vec{K}_{2})+\nonumber\\&&g_{1R,2R}\tilde{C}^{\dagger}_{1,R}(\vec{p}) \tilde{C}_{2,R}(\vec{p})+H.C.
\Big]\nonumber\\&&
H_{L,R}=  \Big[g_{1L,2R}\tilde{C}^{\dagger}_{1,L}(\vec{p})\tilde{C}_{2,R}(\vec{p})+g_{1R,2L}\tilde{C}^{\dagger}_{1,R}(\vec{p})\tilde{C}_{2,L}(\vec{p})+H.C.\Big]\nonumber\\&&
\end{eqnarray}
where the effective tunneling  coefficients are given by:
 \begin{eqnarray}
&&g_{2L,1L}=-\frac{\gamma_{3}}{\gamma_{0}}\Big[\bar{U}^{(2)}_{2,L}(\vec{p}-\vec{K'}_{2})U{(1)}_{1,L}(\vec{p}-\vec{K'}_{1})\phi^{*}_{1}(\vec{p})+\bar{U}^{(1)}_{2,L}(\vec{p}-\vec{K'}_{2})U{(2)}_{1,L}(\vec{p}-\vec{K'}_{1})\phi_{1}(\vec{p})\Big]\nonumber\\&&
g_{2R,1R}=-\frac{\gamma_{3}}{\gamma_{0}}\Big[\bar{U}^{(2)}_{2,R}(\vec{p}-\vec{K}_{2})U{(1)}_{1,R}(\vec{p}-\vec{K}_{1})\phi^{*}_{1}(\vec{p})+\bar{U}^{(1)}_{2,R}(\vec{p}-\vec{K}_{2})U{(2)}_{1,R}(\vec{p}-\vec{K}_{1})\phi_{1}(\vec{p})\Big]\nonumber\\&&
g_{2R,1L}=-\frac{\gamma_{3}}{\gamma_{0}}\Big[\bar{U}^{(2)}_{2,R}(\vec{p}-\vec{K}_{2})U{(1)}_{1,L}(\vec{p}-\vec{K}_{1})\phi^{*}_{1}(\vec{p})+\bar{U}^{(1)}_{2,R}(\vec{p}-\vec{K}_{2})U{(2)}_{1,L}(\vec{p}-\vec{K}_{1})\phi_{1}(\vec{p})\Big]\nonumber\\&&
g_{2L,1R}=-\frac{\gamma_{3}}{\gamma_{0}}\Big[\bar{U}^{(2)}_{2,L}(\vec{p}-\vec{K'}_{2})U{(1)}_{1,R}(\vec{p}-\vec{K}_{1})\phi^{*}_{1}(\vec{p})+\bar{U}^{(1)}_{2,L}we(\vec{p}-\vec{K'}_{2})U{(2)}_{1,R}(\vec{p}-\vec{K}_{1})\phi_{1}(\vec{p})\Big]\nonumber\\&&
\end{eqnarray}
In order to address  the question of flat  bands,   we will solve the model under an  approximations which neglects the  higher order couplings .

 We diagonalize the left $H_{L}$ Hamiltonian and find  two eigenvalues:
\begin{eqnarray}
&&E^{(-)}_{L}=\frac{1}{2}\Big[\epsilon_{1,L}(\vec{p}-\vec{K'}_{1})+\epsilon_{2,L}(\vec{p}-\vec{K'}_{2})-\sqrt{\Big(\epsilon_{1,L}(\vec{p}-\vec{K'}_{1})-\epsilon_{2,L}(\vec{p}-\vec{K'}_{2})\Big)^2+4|g_{1L,2L}|^2}\Big]\nonumber\\&&
E^{(+)}_{L}=\frac{1}{2}\Big[\epsilon_{1,L}(\vec{p}-\vec{K'}_{1})+\epsilon_{2,L}(\vec{p}-\vec{K'}_{2})+\sqrt{\Big(\epsilon_{1,L}(\vec{p}-\vec{K'}_{1})-\epsilon_{2,L}(\vec{p}-\vec{K'}_{2})\Big)^2+4|g_{2L,1L}|^2}\Big]\nonumber\\&&
\end{eqnarray}
We find   the  two operators, $C^{(-)}_{L}$ and $ C^{(+)}_{L}$ using the approximation,  
$E^{(+)}_{L}>>E^{(-)}_{L}$. The approximations is based on  projecting  out   the states $E^{(+)}_{L}$ using  the constraint  $C^{(+)}_{L}\approx 0$  (which is justified when $ E^{(+)}_{L}>>E^{(-)}_{L}$ ).Thus we obtain: 
\begin{eqnarray}
&&\tilde{C}_{2L}(\vec{p})=Z_{2L}C^{(-)}_{L}(\vec{p}),\hspace{0.1in}Z_{2L}=-\frac{\sqrt{(E^{(-)}_{L}-\epsilon_{1,L}(\vec{p}-\vec{K'}_{1}))^2+|g_{1L,2L}|^2}}{(E^{(-)}_{L}-E^{(+)}_{L})}\nonumber\\&&
\tilde{C}_{1L}(\vec{p})=Z_{1L}C^{(-)}_{L}(\vec{p}),\hspace{0.1in} Z_{1L}=-\frac{E^{(-)}_{L}-\epsilon_{1,L}(\vec{p}-\vec{K'}_{1})}{E^{(+)}_{L}-E^{(-)}_{L}}\cdot\frac{\sqrt{(E^{(-)}_{L}-\epsilon_{1,L}(\vec{p}-\vec{K'}_{1}))^2+|g_{1L,2L}|^2}}{g_{1L,2L}}\nonumber\\&&
\end{eqnarray}
We diagonalize the $H_{R}$ Hamiltonian using the approximation based on  projecting  out   the state $E^{(+)}_{L}$ :
\begin{eqnarray}
&&E^{(-)}_{R}=\frac{1}{2}\Big[\epsilon_{1,R}(\vec{p}-\vec{K}_{1})+\epsilon_{2,R}(\vec{p}-\vec{K}_{2})-\sqrt{\Big(\epsilon_{1,R}(\vec{p}-\vec{K}_{1})-\epsilon_{2,R}(\vec{p}-\vec{K}_{2})\Big)^2+4|g_{1R,2R}|^2}\Big]\nonumber\\&&
E^{(+)}_{R}=\frac{1}{2}\Big[\epsilon_{1,R}(\vec{p}-\vec{K}_{1})+\epsilon_{2,R}(\vec{p}-\vec{K}_{2})+\sqrt{\Big(\epsilon_{1,R}(\vec{p}-\vec{K}_{1})-\epsilon_{2,R}(\vec{p}-\vec{K}_{2})\Big)^2+4|g_{1R,2R}|^2}\Big]\nonumber\\&&
\end{eqnarray}
Performing the projection in the right valley gives:
\begin{eqnarray}
&&\tilde{C}_{2R}(\vec{p})=Z_{2R}C^{(-)}_{L}(\vec{p}),\hspace{0.1in}Z_{2R}=-\frac{\sqrt{(E^{(-)}_{R}-\epsilon_{1,R}(\vec{p}-\vec{K}_{1}))^2+|g_{1R,2R}|^2}}{(E^{(-)}_{R}-E^{(+)}_{R})}\nonumber\\&&
\tilde{C}_{1R}(\vec{p})=Z_{1R}C^{(-)}_{R}(\vec{p}),\hspace{0.1in} Z_{1R}=-\frac{E^{(-)}_{R}-\epsilon_{1,R}(\vec{p}-\vec{K}_{1})}{E^{(+)}_{R}-E^{(-)}_{R}}\cdot\frac{\sqrt{(E^{(-)}_{R}-\epsilon_{1,R}(\vec{p}-\vec{K}_{1}))^2+|g_{1R,2R}|^2}}{g_{1R,2R}}\nonumber\\&&
\end{eqnarray}
The effective tunneling Hamiltonian which includes the coupling between the two valleys is:
\begin{eqnarray}
&&H_{eff.}=\sum_{\vec{p}}\nonumber\\&&C^{(-)\dagger}_{L}(\vec{p})C^{(-)}_{L}(\vec{p})E^{(-)}_{L}(\vec{p})+C^{(-)\dagger}_{R}(\vec{p})C^{(-)}_{R}(\vec{p})E^{(-)}_{R}(\vec{p})+G\cdot C^{(-)\dagger}_{L}(\vec{p})C^{(-)}_{R}(\vec{p})+G^{*}\cdot C^{(-)\dagger}_{R}(\vec{p})C^{(-)}_{R}(\vec{p})\nonumber\\&&
G=g_{1L,2R}Z^{*}_{1L}Z_{2R}+(g_{1L,2R})^{*}Z^{*}_{2L}Z_{1R})\nonumber\\&&
\end{eqnarray}
The lowest eigenvalue  of the effective Hamiltonian will give the band  $E_{0}$:
\begin{equation}
E_{0}\frac{1}{2}\Big[E^{(-)}_{L}(\vec{p})+E^{(-)}_{R}(\vec{p})-\sqrt{\Big(E^{(-)}_{L}(\vec{p})-E^{(-)}_{R}(\vec{p})\Big)^2+4|G|^{2}}\Big]
\label{eigenvalue}
\end{equation}
The lowest energy band $E_{0}$ is   flat $E_{flat}=E_{0}$ and justifies the name  when $ \theta\approx\theta(n)_{magic}$

\vspace{0.2in }

\textbf{V-Superconductivity induced by the rotated  layer $ i=2$}

\vspace{0.2in }

Here we will use the spinor representation to demonstrate  that the rotation by angle   $\theta$  affects the electron-electron  interactions. For simplicity we consider a repulsive Hubbard interaction.   Due to the spinor rotation    the Hubbard interaction  becomes attractive   in the  $y$ direction  and  periodic   in the $x$  direction .Effectively, this is described     as a set  of  one dimensional superconducting   wires separated by metallic regions.
This result is  additive to the  attractive  interactions mediated by the phonons.

  The interaction in layer $i=2$ is controlled by   two fields , $\Psi_{2}(\vec{R'})$ and $\Psi_{2}(\vec{R'}+\delta_{\vec{r'}})$: 
\begin{eqnarray}
&&\Psi_{2}(\vec{R'})=\sum_{\vec{p}}\Big[\tilde{C}_{2,L}(\vec{p})U_{2,L}(\vec{p}-\vec{K'}_{2})e^{i\vec{p}[-\theta]\cdot\vec{R}}+\tilde{C}_{2,R}(\vec{p})U_{2,R}(\vec{p}-\vec{K}_{2})e^{i\vec{p}[-\theta]\cdot\vec{R}}\Big]\nonumber\\&&
=\sum_{0<p_{x}<\Lambda}\sum_{-\Lambda <p_{y}<\Lambda}\Big[\tilde{C}_{2,L}(-\vec{p})U_{2,L}(\vec{p}-\vec{K'}_{2})e^{i(-p_{x}[-\theta]\cdot R_{x}+p_{y}[-\theta])\cdot R_{y}}+\tilde{C}_{2,R}(\vec{p})U_{2,R}(\vec{p}-\vec{K}_{2})e^{i\vec{p}[-\theta]\cdot\vec{R}}\Big]
\nonumber\\&&
\Psi_{2}(\vec{R'}+\delta_{\vec{r'}})\approx\sum_{0<p_{x}<\Lambda}\sum_{-\Lambda <p_{y}<\Lambda}\Big[\tilde{C}_{2,L}(-\vec{p})U_{2,L}(-\vec{K'}_{2})e^{i(-p_{x}[-\theta]\cdot R_{x}+p_{y}[-\theta])\cdot R_{y})+i(-p_{x}[-\theta]\cdot \delta_{r_{x}}+p_{y}[-\theta]\cdot \delta_{r_{y}})}\nonumber\\&&+\tilde{C}_{2,R}(\vec{p})U_{2,R}(-\vec{K}_{2})e^{i\vec{p}[-\theta]\cdot(\vec{R}+\delta_{\vec{r}})}\Big]\nonumber\\&&
\Psi_{2}(\vec{R'})\approx\sum_{0<p_{x}<\Lambda}\sum_{-\Lambda <p_{y}<\Lambda}\Big[\tilde{C}_{2,L}(-\vec{p})U_{2,L}(-\vec{K'}_{2})e^{i(-p_{x}[-\theta]\cdot R_{x}+p_{y}[-\theta])\cdot R_{y})}+\nonumber\\&&\tilde{C}_{2,R}(\vec{p})U_{2,R}(-\vec{K}_{2})e^{i\vec{p}[-\theta]\cdot\vec{R}}\Big]
\nonumber\\&&
\end{eqnarray}
where  the spinors are given by: $U_{2,L}(\vec{p}-\vec{K'}_{2})\approx U_{2,L}(-\vec{K'}_{2})= \frac{1}{\sqrt{2}}\Big[1,ie^{i\theta}e^{-i\vec{K'}_{2}}\Big]$,

\noindent
$U_{2,R}(\vec{p}-\vec{K}_{2})\approx U_{2,R}(-\vec{K}_{2})= \frac{1}{\sqrt{2}}\Big[1,ie^{i\theta}e^{-\vec{K}_{2}}\Big]$.

\noindent
 We observe that the spinor depends on the a rotated   angle $\theta$.

\noindent
The Hubbard interaction in layer $i=2$ is:
\begin{eqnarray}
&&H_{U}=\sum_{\vec{R}}U n_{\uparrow}(\vec{R'})n_{\downarrow}(\vec{R'})+\sum_{\vec{R}}\sum_{r=1,2,3}U n_{\uparrow}(\vec{R'}+\delta_{\vec{r'}})n_{\downarrow}(\vec{R'}+\delta_{\vec{r'}})\nonumber\\&&
\end{eqnarray}
We will use   the summation over  the  twisted vector $\vec{\delta}_{r'}$ introduced in equation $(8)$ which relates the rotated  momentum to the $Moire$  reciprocal lattice (see Eq.$(6)$ ) :
\begin{eqnarray}
&&\sum_{r=1,2,3}e^{-i(\vec{p}_{1}[-\theta]+\vec{p}_{2}[-\theta]+\vec{p}_{3}[-\theta]+\vec{p}_{4}[-\theta])\cdot\vec{\delta}_{r}}\delta_({\vec{p}_{1}[-\theta]+\vec{p}_{2}[-\theta]+\vec{p}_{3}[-\theta]+\vec{p}_{4}[-\theta])_{x},g_{x}[\theta]}\nonumber\\&&\cdot\delta_({\vec{p}_{1}[-\theta]+\vec{p}_{2}[-\theta]+\vec{p}_{3}[-\theta]+\vec{p}_{4}[-\theta])_{y},g_{y}[\theta]}\nonumber\\&&
=e^{-i (g_{x}[\theta]+g_{y}[\theta])}\Big [1+2e^{i(\frac{3}{2}g_{x}[\theta]+\frac{3}{2}g_{y}[\theta])}Cos\Big[\frac{\sqrt{3}}{2}(g_{x}[\theta]-g_{y}[\theta])\Big]\nonumber\\&&
\end{eqnarray}
Next,   in the Hubbard  interaction we substitute the fields $\Psi_{2,\sigma}(\vec{R'})$ and $\Psi_{2,\sigma}(\vec{R'}+\delta_{\vec{r'}})$  given in Eq.$(33)$ including the spin dependence. We notice that  the intervalleys    depend on the rotated  angles . If  we  use  a long range Coulomb potential we can use the same strategy as used for the Hubbard model , by identifying  directions where  we have an  attractive potential.Those terms  are given by $\tilde{C}^{\dagger}_{2,L,\uparrow}(\vec{p}_{1})\tilde{C}^{\dagger}_{2,L,\downarrow}(\vec{p}_{2})\tilde{C}_{2,R,\downarrow}(\vec{p}_{3})\tilde{C}_{2,R,\uparrow}(\vec{p}_{4})$ and $\tilde{C}^{\dagger}_{2,R,\uparrow}(\vec{p}_{1})\tilde{C}^{\dagger}_{2,R,\downarrow}(\vec{p}_{2})\tilde{C}_{2,L,\downarrow}(\vec{p}_{3})\tilde{C}_{2,L,\uparrow}(\vec{p}_{4})$.
We introduce   the short notation   :
\begin{eqnarray}
&&\sum^{p_{x}}=\sum_{0<p_{2x}<\Lambda}\sum_{0<p_{3x}<\Lambda}\sum_{0<p_{4x}<\Lambda}\nonumber\\&& 
\sum^{p_{y}}=\sum_{-\Lambda <p_{2y}<\Lambda}\sum_{-\Lambda <p_{3y}<\Lambda}\sum_{-\Lambda <p_{4y}<\Lambda}\nonumber\\&&
\end{eqnarray}
Using the periodicity with respect to  the $Moire$ reciprocal  lattice   $\tilde{C}_{2,L}(\vec{p}+\vec{g}[\theta])=\tilde{C}_{2,L}(\vec{p})$, $\tilde{C}_{2,R}(\vec{p}+\vec{g}[\theta])=\tilde{C}_{2,R}(\vec{p})$ we obtain the following  form of the Hubbard interaction:
\begin{eqnarray}
&&H_{U}=\sum^{p_{y}}
\sum^{p_{x}}\nonumber\\&&\Big[\frac{U}{4}\Big(e^{-i 4\theta}+e^{-i( 4\theta+g_{x}[\theta]+g_{y}[\theta])}+2e^{-i (4\theta-\frac{1}{2}g_{x}[\theta]-\frac{1}{2}g_{y}[\theta])}Cos[\frac{\sqrt{3}}{2}(g_{x}[\theta]-g_{y}[\theta])\Big)\nonumber\\&&\tilde{C}^{\dagger}_{2,R,\uparrow}(-p_{2x}-p_{3x}-p_{4x},-p_{2y}-p_{3y}-p_{4y})\tilde{C}^{\dagger}_{2,R,\downarrow}(p_{2x},p_{2y})\tilde{C}_{2,L,\downarrow}(-p_{3x},p_{3y})\tilde{C}_{2,L,\uparrow}(-p_{4x},p_{4y})\nonumber\\&&+\frac{U}{4}\Big(e^{i 4\theta}+e^{i( 4\theta+g_{x}[\theta]+g_{y}[\theta])}+2e^{i (4\theta-\frac{1}{2}g_{x}[\theta]-\frac{1}{2}g_{y}[\theta])\Big)}Cos[\frac{\sqrt{3}}{2}(g_{x}[\theta]-g_{y}[\theta])\Big)\cdot\nonumber\\&&\tilde{C}^{\dagger}_{2,L,\uparrow}(p_{2x}+p_{3x}+p_{4x},-p_{2y}-p_{3x}-p_{4x})\tilde{C}^{\dagger}_{2,L,\downarrow}(-p_{2x},p_{2y})\tilde{C}_{2,R,\downarrow}(p_{3x},p_{3y})\tilde{C}_{2,R,\uparrow}(p_{4x},p_{4y})\nonumber\\&&
\end{eqnarray}
At certain angles, the imaginary part of the effective Hubbard potential vanishes and the real part is negative:
\begin{eqnarray}
&&Im.\Big[\frac{U}{4}\Big(e^{-i 4\theta}+e^{-i( 4\theta+g_{x}[\theta]+g_{y}[\theta])}+2e^{-i (4\theta-\frac{1}{2}g_{x}[\theta]-\frac{1}{2}g_{y}[\theta])}Cos[\frac{\sqrt{3}}{2}(g_{x}[\theta]-g_{y}[\theta])\Big)\Big]=0\nonumber\\&&
U[\bar{\theta}]=\frac{U}{2}\Big(Cos[4\theta]+Cos[4\theta+g_{x}[\theta]+g_{y}[\theta]]+2Cos[(4\theta-\frac{1}{2}g_{x}[\theta]-\frac{1}{2}g_{y}[\theta])Cos[\frac{\sqrt{3}}{2}(g_{x}[\theta]-g_{y}[\theta])]\nonumber\\&&\Big)<0\nonumber\\&&
\end{eqnarray}
We define :

\noindent
$U[\bar{\theta}]=\frac{U}{2}\Big(Cos[4\theta]+Cos[4\theta+g_{x}[\theta]+g_{y}[\theta]]+2Cos[(4\theta-\frac{1}{2}g_{x}[\theta]-\frac{1}{2}g_{y}[\theta])Cos[\frac{\sqrt{3}}{2}(g_{y}[\theta]-g_{x}[\theta])]\Big)$. 
\noindent
We observe in Figure $ 1$ that  the  interaction  term $U[\bar{\theta}]$ is  attractive for   certain angles $\theta=s$.
\begin{eqnarray}
&&H_{U}\approx\sum^{p_{y}}\sum^{p_{x}}\nonumber\\&&\Big[-|U[\bar{\theta}]|  
\Big(\tilde{C}^{\dagger}_{2,R,\uparrow}(-p_{2x}-p_{3x}-p_{4x},-p_{2y}-p_{3x}-p_{4x})\tilde{C}^{\dagger}_{2,R,\downarrow}(p_{2x},p_{2y})\tilde{C}_{2,L,\downarrow}(-p_{3x},p_{3y})\tilde{C}_{2,L,\uparrow}(-p_{4x},p_{4y})\nonumber\\&&+\tilde{C}^{\dagger}_{2,L,\uparrow}(p_{2x}+p_{3x}+p_{4x},-p_{2y}-p_{3y}-p_{4y})\tilde{C}^{\dagger}_{2,L,\downarrow}(-p_{2x},p_{2y})\tilde{C}_{2,R,\downarrow}(p_{3x},p_{3y})\tilde{C}_{2,R,\uparrow}(p_{4x},p_{4y})\Big)\Big]\nonumber\\&&
\end{eqnarray}
This allows us  to write a one dimensional  pairing Hamiltonian at a fixed linear  momentum $ p_{x}=0$ ( this corresponds  to $\vec{K'}_{2,x}= \vec{K}_{2,x}$) :
\begin{eqnarray}
&&H_{U}=-|U[\bar{\theta}]|\int\,dy\Big[\tilde{C}^{\dagger}_{2,R,\uparrow}(p_{x}=0,y)\tilde{C}^{\dagger}_{2,R,\downarrow}(p_{x}=0,y)\tilde{C}_{2,L,\downarrow}(p_{x}=0,y)\tilde{C}_{2,L,\uparrow}(p_{x}=0,y)+\nonumber\\&&\tilde{C}^{\dagger}_{2,L,\uparrow}(p_{x}=0,y)\tilde{C}^{\dagger}_{2,L,\downarrow}(p_{x}=0,y)\tilde{C}_{2,R,\downarrow}(p_{x}=0,y)\tilde{C}_{2,R,\uparrow}(p_{x}=0,y)\Big]\nonumber\\&&
\end{eqnarray}  
Following \cite{Weinberg} we use  the Lagrangian  representation  and  perform a saddle 
point computation:
\begin{eqnarray}
&&L=L_{0}+L_{int.}\nonumber\\&&
L_{int.}=|U[\bar{\theta}]|\int\,dy\Big[\tilde{C}^{\dagger}_{2,R,\uparrow}(p_{x}=0,y)\tilde{C}^{\dagger}_{2,R,\downarrow}(p_{x}=0,y)\tilde{C}_{2,L,\downarrow}(p_{x}=0,y)\tilde{C}_{2,L,\uparrow}(p_{x}=0,y)+\nonumber\\&&\tilde{C}^{\dagger}_{2,L,\uparrow}(p_{x}=0,y)\tilde{C}^{\dagger}_{2,L,\downarrow}(p_{x}=0,y)\tilde{C}_{2,R,\downarrow}(p_{x}=0,y)\tilde{C}_{2,R,\uparrow}(p_{x}=0,y)\Big]\nonumber\\&&
\end{eqnarray}
We use  the  Hubbard  Stratonovici  fields $\Delta$,$\Delta^{*}$, $D$ and $D^{*}$. This is done by replacing:

\noindent
$ \tilde{C}^{\dagger}_{2,R,\uparrow}(p_{x}=0,y)\tilde{C}^{\dagger}_{2,R,\downarrow}(p_{x}=0,y)\tilde{C}_{2,L,\downarrow}(p_{x}=0,y)\tilde{C}_{2,L,\uparrow}(p_{x}=0,y)=\frac{1}{4}\Big(\tilde{C}^{\dagger}_{2,R,\uparrow}(p_{x}=0,y)\tilde{C}^{\dagger}_{2,R,\downarrow}(p_{x}=0,y)+\tilde{C}_{2,L,\downarrow}(p_{x}=0,y)\tilde{C}_{2,L,\uparrow}(p_{x}=0,y)\Big)^2-\frac{1}{4}\Big(\tilde{C}^{\dagger}_{2,R,\uparrow}(p_{x}=0,y)\tilde{C}^{\dagger}_{2,R,\downarrow}(p_{x}=0,y)-\tilde{C}_{2,L,\uparrow}(p_{x}=0,y)\tilde{C}_{2,L,\downarrow}(p_{x}=0,y)^2$

\noindent
and further decouplingthe two body interaction   by  a Gaussian  integration \cite{Weinberg}.
\begin{eqnarray}
&&\int\,dt L_{int.}(t)=\int\,dt\int\,dy\Big[-\frac{\Delta(p_{x}=0,y;t)\Delta^{*}(p_{x}=0,y;t)}{4|U[\bar{\theta}]|}-\frac{D(p_{x}=0,y;t) D^{*}(p_{x}=0,y;t)}{4|U[\bar{\theta}]|}\nonumber\\&&
+\tilde{C}^{\dagger}_{2,L,\uparrow}(p_{x}=0,y;t)\tilde{C}^{\dagger}_{2,L,\downarrow}(p_{x}=0,y;t)\Delta(p_{x}=0,y;t)+\nonumber\\&&\tilde{C}_{2,R,\downarrow}(p_{x}=0,y)\tilde{C}_{2,R,\uparrow}(p_{x}=0,y;t)\Delta^{*}(p_{x}=0,y;t)+\nonumber\\&&
\tilde{C}^{\dagger}_{2,R,\uparrow}(p_{x}=0,y;t)\tilde{C}^{\dagger}_{2,R,\downarrow}(p_{x}=0,y)D(p_{x}=0,y;t)+\nonumber\\&&\tilde{C}_{2,L,\downarrow}(p_{x}=0,y;t)\tilde{C}_{2,L,\uparrow}(p_{x}=0,y;t)D^{*}(p_{x}=0,y;t)\Big]\nonumber\\&&
\end{eqnarray}
The variation with respect to  the fields  $\Delta$,$\Delta^{*}$, $D$ and $D^{*}$ gives the
equations :
\begin{eqnarray}
&&\Delta^{*}(p_{x}=0,y;t)=4\langle \tilde{C}^{\dagger}_{2,L,\uparrow}(p_{x}=0,y;t)\tilde{C}^{\dagger}_{2,L,\downarrow}(p_{x}=0,y;t)\rangle,\nonumber\\&&\Delta(p_{x}=0,y;t)=4\langle \tilde{C}_{2,R,\uparrow}(p_{x}=0,y;t)\tilde{C}_{2,R,\downarrow}(p_{x}=0,y;t)\rangle;
\nonumber\\&&D(p_{x}=0,y;t)=4\langle \tilde{C}_{2,L,\uparrow}(p_{x}=0,y;t)\tilde{C}_{2,L,\downarrow}(p_{x}=0,y;t)\rangle,\nonumber\\&&D^{*}(p_{x}=0,y;t)=4\langle \tilde{C}{\dagger}_{2,R,\uparrow}(p_{x}=0,y;t)\tilde{C}^{\dagger}_{2,R,\downarrow}(p_{x}=0,y;t)\rangle;\nonumber\\&&
\end{eqnarray}
From these equations we obtain :

\noindent
$ \langle \tilde{C}^{\dagger}_{2,L,\uparrow}(p_{x}=0,y;t)\tilde{C}^{\dagger}_{2,L,\downarrow}(p_{x}=0,y;t)\rangle=\Big(\langle \tilde{C}_{2,L,\uparrow}(p_{x}=0,y;t)\tilde{C}_{2,L,\downarrow}(p_{x}=0,y;t)\rangle\Big)^{*}$.
This allows us to reduce the four fields  to only two:$D=\Delta=\chi(p_{x}=0,y)$, $\Delta^{*}=D^{*}=\chi(p_{x}=0,y)^{*}$.Using   this saddle point, we simplify  the interaction term:
\begin{eqnarray}
&&\int\,dt L_{int.}(t)=\int\,dt\int\,dy\Big[-\frac{\chi(p_{x}=0,y;t)\chi^{*}(p_{x}=0,y;t)}{2|U[\bar{\theta}]|}\nonumber\\&&
+\Big(\tilde{C}^{\dagger}_{2,L,\uparrow}(p_{x}=0,y;t)\tilde{C}^{\dagger}_{2,L,\downarrow}(p_{x}=0,y;t)+\tilde{C}^{\dagger}_{2,R,\uparrow}(p_{x}=0,y)\tilde{C}^{\dagger}_{2,R,\downarrow}(p_{x}=0,y;t)\Big)\chi(p_{x}=0,y;t)+\nonumber\\&&
\chi^{*}(p_{x}=0,y;t)\Big(\tilde{C}_{2,R,\downarrow}(p_{x}=0,y;t)\tilde{C}_{2,R,\uparrow}(p_{x}=0,y)+\tilde{C}_{2,L,\downarrow}(p_{x}=0,y;t)\tilde{C}_{2,L,\uparrow}(p_{x}=0,y;t)\Big)\Big]\nonumber\\&&
\end{eqnarray}
This gives a   one-dimensional superconducting order  parameter  periodic in $x$ with periodicity $ K_{1,x}$.
\begin{equation}
\chi(p_{x}=0,y;t)=Cos[K_{1,x}x]\chi_{0}(T)
\label{order}
\end{equation}
To compute the critical temperature $T=T_{c}$, we perform a variation with  respect to  $\chi$ . Performing  the computation in   the  Euclidean form, we obtain:
\begin{eqnarray}
&&L=-\frac{|\chi(p_{x}=0)|^{2}}{2|U[\bar{\theta}]|}-\frac{T}{L}\sum_{\omega_{n}}\sum_{p}Log\Big[(i\omega_{n})^{4}-E^4-|\chi|^4\Big];\nonumber\\&&
Log\Big[(i\omega_{n})^{4}-E^4-|\chi|^4\Big]=Log\Big[\Big(i\omega_n-\sqrt{E^{2}+|\chi|^{2}})\Big)\Big(i\omega_n+\sqrt{E^{2}+|\chi|^{2}}\Big)\Big(i\omega_n-i\sqrt{E^2+|\chi|^{2}}\Big))\nonumber\\&&\cdot\Big(i\omega_n+i\sqrt{E^2+|\chi|^{2}}\Big)\Big]\nonumber\\&&
\end{eqnarray}
We obtain the equation:
\begin{eqnarray}
&&\frac{1}{2|U[\bar{\theta}]|}=\frac{1}{L}\sum_{p}\Big[\frac{1}{E(p)}tanh(\frac{E(p)}{2T})+\frac{Sin[\frac{E(p)}{T}]}{1+Cos[\frac{E(p)}{T}]}\Big]\nonumber\\&&
\end {eqnarray}

with $E(\vec{p})=\epsilon(\vec{p}) -\mu<\Lambda$

 The elctronic dispersion is not known.The  flat band allows us  to replace $E(p)=\epsilon(\vec{p})_{max} -\mu\approx 0$ and  superconductivity is destroyed .
 The critical temperature for  limit  $\mu\approx 0$  gives $T_{ c}\approx \frac{L}{\epsilon(\vec{p})_{max}}|U[\bar{\theta}]|$.

\vspace{0.2 in}

\textbf{VI-Conclusions}

\noindent
To conclude, we have shown that in graphene  lower rotation  induces an attractive interaction for some directions and certain valley components. This results   in  a   one-dimensional superconducting parameter  periodic in $x$ with periodicity $ K_{1,x}$ .We want to mention that  alternative explanations have  been proposed .These explanations are  complementary    to  the model proposed  here   and can give rise to two dimensional superconductivity.

\noindent
These results have been obtained under the following conditions:

\noindent
 a)A superlattice  with   basis vectors $\vec{t}_{1}=i\vec{a}_{1}+(i+1)\vec{a}_{2}$;
$\vec{t}_{2}=-(i+1)\vec{a}_{1}+(2i+1)\vec{a}_{2}$ is formed at magic angles $cos[\theta_{i}]=\frac{3i^2+3i+0.5}{3i^2+3i+1}$,i=0,1,2...

 \noindent
b)A uniform rotation in real space by an angle $\theta$ is equivalent to a rotation $-\theta$  of  the  momentum  space.

\noindent
c)  The discrete summation with respect to  the lattice points relates the momentum to the reciprocal lattice vectors $(G^{(1)}_{x}+G^{(2)}_{x})Cos[\theta]$ and $(G^{(1)}_{x}+G^{(2)}_{x})Sin[\theta]$  which at the magic angles become the $Moire$ reciprocal lattice .

\begin{figure}
\begin{center}
\includegraphics[width=5. in]{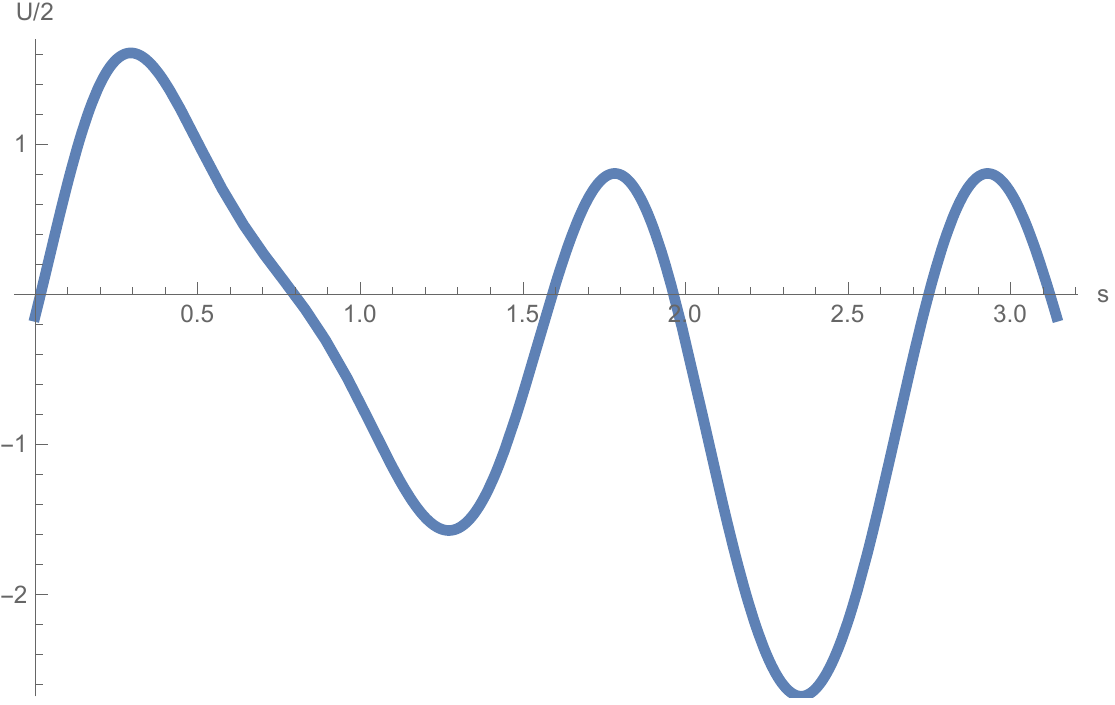}
\end{center}
\caption{The possible attractive potential, $ U[\bar{\theta}]=\frac{U}{2}\Big(Cos[4\theta]+Cos[4\theta+g_{x}[\theta]+g_{y}[\theta]]+2Cos[(4\theta-\frac{1}{2}g_{x}[\theta]-\frac{1}{2}g_{y}[\theta])Cos[\frac{\sqrt{3}}{2}(g_{y}[\theta]-g_{x}[\theta])]\Big)$. }
\end{figure}

\end{document}